\documentclass[sigconf]{acmart}
\usepackage{graphicx}
\usepackage{array}
\usepackage{multirow}

\usepackage{import}
\usepackage{longtable}

\acmConference[Workshop "Dreaming Disability Justice in HCI"]{}{April 30--May 05, 2022}{New Orleans (Virtual Workshop)}
  
\acmBooktitle{Position Paper at the Workshop "Dreaming Disability Justice in HCI" at the CHI Conference on Human Factors in Computing Systems (CHI'22), April 30--May 05, 2022, New Orleans (Virtual Workshop),LA.} 
\acmDOI{}

\AtBeginDocument{%
  \providecommand\BibTeX{{%
    \normalfont B\kern-0.5em{\scshape i\kern-0.25em b}\kern-0.8em\TeX}}}

\begin{document}

\title[]{Justice in interaction design: preventing manipulation in interfaces}
\begin{abstract}
 Designers incorporate values in the design process that raise risks for vulnerable groups. Persuasion in user interfaces can quickly turn into manipulation and become potentially harmful for those groups in the realm of intellectual disabilities, class, or health, requiring proactive responsibility approaches in design. Here we introduce the Capability Sensitive Design Approach and explain how it can be used proactively to inform designers' decisions when it comes to evaluating justice in their designs preventing the risk of manipulation.  
\end{abstract}

\author{Lorena Sánchez Chamorro}
\email{lorena.sanchezchamorro@uni.lu}
\affiliation{%
  \institution{University of Luxembourg}
  \country{Luxembourg}
}
\author{Kerstin Bongard-Blanchy}
\email{kerstin.bongardblanchy@uni.lu}
\affiliation{%
  \institution{University of Luxembourg}
  \country{Luxembourg}
}
\author{Vincent Koenig}
\email{vincent.koenig@uni.lu}
\affiliation{%
  \institution{University of Luxembourg}
  \country{Luxembourg}
}
\renewcommand{\shortauthors}{Sánchez Chamorro et al.}


\maketitle

\section{Designers as 'mediators of ethics'}
Designers become "mediators of ethics" \cite{gray_ethical_2019} when they incorporate their own values in the design process and outcome  \cite{winner_artifacts_1980,gillespie_politics_2010,harrington_eliciting_2021,bardzell_reading_2014}. This gives rise to several risks for vulnerable groups such as hindering them to accomplish their interaction goal \cite{holmes_cycle_2020,bardzell_reading_2014,carter_critical_2004}, developing misrepresentations and perpetuating stereotypes \cite{constanza-chock_design_2020}. Critical design scholars have discussed the ideal momentum to apply ethical considerations in the design process \cite{bardzell_reading_2014}. Although evaluating the consequences of certain design choices can help scrutinise design flaws, it may fail to unveil the problems inherent to the design process, such as limited use of personas and prioritising inappropriate stakeholders to contest power. 

Although Value-Conscious approaches have tried to integrate ethics in design, they fail to proactively recognise exclusion and connect with the practitioners' mindset \cite{gray_dark_2018}. As Jacobs \cite{jacobs_why_2021} explains, these frameworks need to be loaded with values that designers as mediators of ethics \cite{gray_ethical_2019} use discretionary without a guarantee of justice by default. Berdichevsky et al. \cite{berdichevsky_toward_1999} suggested that designers are responsible for the consequences of their design choices only if those are reasonably predictable. As Gray \cite{gray_dark_2018} argues, acknowledging the designer's responsibility is necessary to engage in ethical designs. Nevertheless, to disentangle and avoid those reasonably predictable consequences, it is necessary to include proactive approaches in the design process that lead to ethical designs. 

Interaction design is not exempt from impacting vulnerable groups. The term \textit{'vulnerable groups'} refers to collectives in a less privileged position alongside different axes: non-healthy vs healthy, LGTBI vs heterosexual, or women vs men \cite{constanza-chock_design_2020}. In the same way that a lock "disaffords" people to open a door if they do not have the key, some designs "disafford" users, preventing them from equally benefiting from the design \cite{wittkower_principles_2016, constanza-chock_design_2020}. 
 
\section{When persuasion becomes manipulation for vulnerable groups}
 
The problem of a lack of a definition of justice in the design process becomes very present in the realm of persuasive design. Although persuasion as a way of influencing behaviour is commonly used \cite{jacobs_two_2020, fogg_behavior_2009}, there are no unequivocal boundaries to distinguish between manipulation and persuasion. While manipulation is usually hidden \cite{susser_online_2018}, persuasion should be transparent for the user \cite{thaler_chapter_2013}. 

\begin{figure*}[h]
    \centering
    \includegraphics[width=12cm]{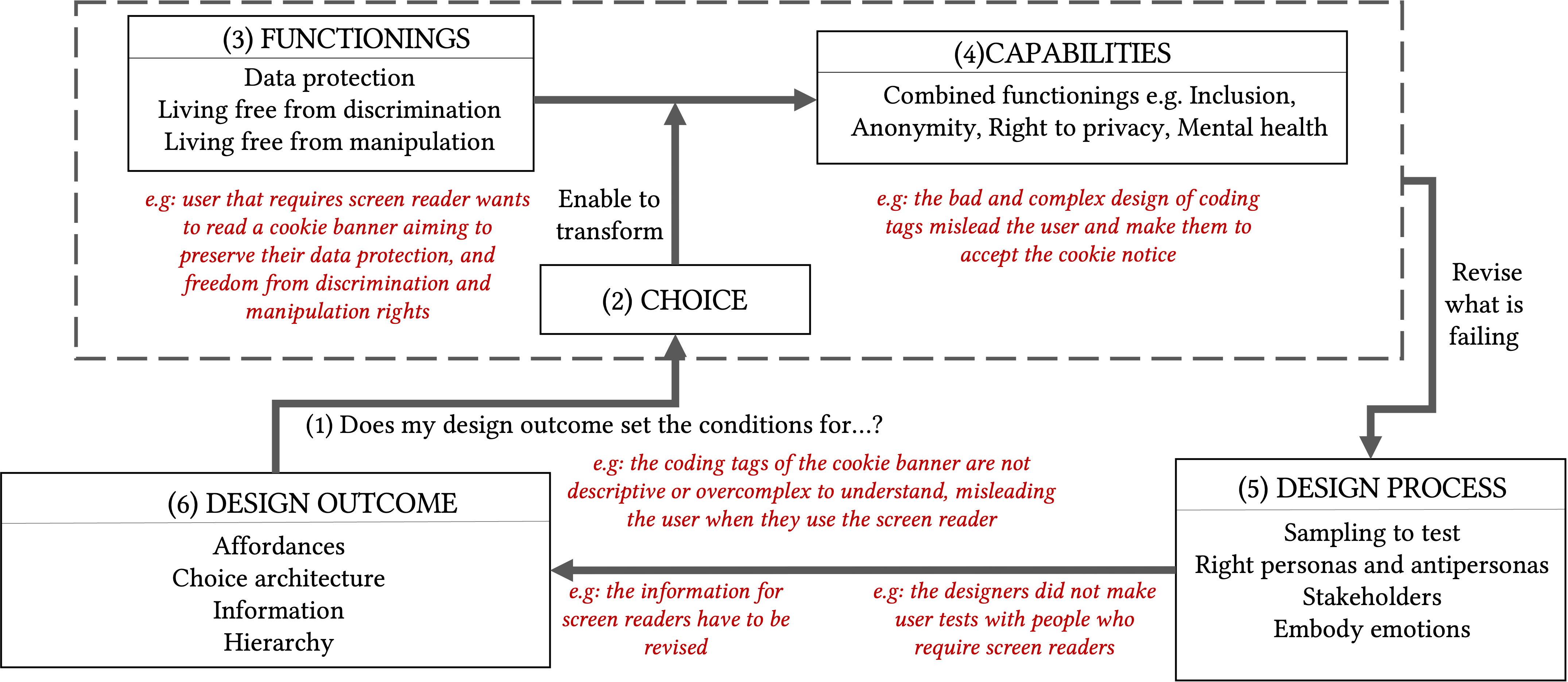}
    \caption{A conceptualisation of the Capability Sensitive Design Approach within the design process}
    \label{fig:galaxy}
\end{figure*}

The line between manipulation and persuasion may be different for specific vulnerable groups, being the strategies to develop manipulative interfaces more adverse for them \cite{gray_dark_2018}. The widespread use of persuasive design in mobile phone applications \cite{di_geronimo_ui_2020-1}, videogames \cite{denoo_dark_2021} and social media can cross the line between manipulation and persuasion. Some collectives may find persuasive elements less transparent depending on users' intellectual abilities, education levels, or mental disorders. Therefore, if designers disregard these structural factors, persuasion can quickly become manipulation. 

Gak et al. \cite{gak_distressing_2021} and Jacobs \cite{jacobs_two_2020} discussed how the use of persuasive design in ads on Instagram and calorie counting applications could be especially harmful to people with eating disorders. Manipulative designs use addictive components, being more damaging for lower classes that are more likely to develop an internet addiction \cite{lee_socioeconomic_2015}. This addiction also relates to anxiety and depression tendencies \cite{choi_dysfunctional_2014}, giving rise to an intersectional risk between class and health. Jenaro et al. \cite{jenaro_internet_2018} also found that adults with intellectual disabilities use the Internet mainly for playing and meeting people on the mobile phone applications, where there is a high level of persuasive designs present \cite{di_geronimo_ui_2020-1}. These persuasive techniques can be manipulative when users do not transparently perceive the context. Certain groups (e.g., children or people with intellectual disabilities) cannot understand the business model behind "pay to win" in video games. Therefore, while for adults without intellectual disabilities being nudged into paying can be a persuasive technique, this may fall into the manipulation realm for other collectives. 

\section{Defining Justice through the Capability Sensitive Design Approach}
 
We propose the Capability Sensitive Design \cite{slavova_towards_2013, jacobs_capability_2020} (from now on "CSD approach") as an approach in the interaction design process that allows recognising structural problems of persuasion on vulnerable groups. Defining the different 
needs that users expect to 
cover when interacting with the interface, will help designers inform their decisions during the design process from a proactive perspective. 

The CSD approach considers justice a matter of choice. It is about the possibility of choosing what a person values in terms of doings or beings\cite{jacobs_capability_2020, sen_freedom_1999} (\textit{"functionings"}) and making them happen. Only when every user is allowed to choose their own set of \textit{"functionings"} and translate them into \textit{"capabilities"} it is possible to talk about equal treatment.
To translate \textit{"functionings"} into \textit{"capabilities"}, design needs to provide the proper conditions.
As an example, the functioning \textit{"having a sexual orientation"} will not be transformed into the capability of \textit{"expressing sexual orientation"} if a web form does not recognise all types of sexual orientation\cite{jacobs_capability_2020}.

Therefore, we consider the CSD approach as a crucial piece towards reflectivity in the design process that can help conceptualise justice in interaction design to address the boundaries of manipulation. 
CSD stimulates reflection on the set of \textit{"functionings"} that need to be translated into \textit{"capabilities"} through the design process. As shown in Figure 1, (1) a designer should question to what extent the design is not allowing the specific user (2) to choose (4) the capabilities they seek because (3) the functionings cannot be transformed.  
In that case, (5) the design process needs to be evaluated, analysing in which step vulnerable groups are misrepresented, and consequently, (6) the design outcome will change. 
\section{Acknowledgements}
 This work has been supported by the Luxembourg National Research Fund (FNR) -- IS/14717072 ``Deceptive Patterns Online (Decepticon)’’.

\bibliographystyle{ACM-Reference-Format}
\bibliography{references.bib}

\end{document}